# The ECLAIRs GRB-trigger telescope on-board the future mission SVOM

**Stéphane Schanne**[1], **Bertrand Cordier**
*CEA Saclay, IRFU/Service d'Astrophysique, 91191 Gif sur Yvette, France*

**Jean-Luc Atteia, Olivier Godet**
*IRAP, 14 avenue E. Belin, 31400 Toulouse, France*

**Cyril Lachaud**
*APC (UMR 7164), 10 rue Alice Domon et Léonie Duquet, 75205 Paris, France*

**Karine Mercier**
*CNES, 18 avenue E. Belin, 31401 Toulouse, France*

The Space-based multi-band astronomical Variable Objects Monitor (SVOM) is an approved satellite mission for Gamma-Ray Burst (GRB) studies, developed in cooperation between the Chinese National Space Agency (CNSA), the Chinese Academy of Sciences (CAS), the French Space Agency (CNES) and French laboratories. SVOM entered Phase B in 2014 and is scheduled for launch in 2021. SVOM will provide fast and accurate GRB localizations, and determine the temporal and spectral properties of the GRB emission, thanks to a set of 4 on-board instruments. The trigger system of the coded-mask telescope ECLAIRs images the sky in the 4-120 keV energy range, in order to detect and localize GRBs in its 2 sr-wide field of view. The low-energy threshold of ECLAIRs is well suited for the detection of highly redshifted GRB. The high-energy coverage is extended up to 5 MeV thanks to the non-imaging gamma-ray spectrometer GRM. GRB alerts are sent in real-time to the ground observers community, and a spacecraft slew is performed in order to place the GRB within the field of view of the soft X-ray telescope MXT and the visible-band telescope VT, to refine the GRB position and study its early afterglow. The ground-based robotic telescopes GFTs and the wide angle cameras GWAC complement the on-board instruments. In this paper we present the ECLAIRs soft gamma-ray imager which will provide the GRB triggers on-board SVOM.

[1] Speaker, E-mail: stephane.schanne@cea.fr

**1. Introduction to SVOM**

A Gamma-Ray Burst (GRB) is a cosmic event believed to accompany the formation of a stellar-mass black hole or a magnetar at cosmological distances. A gamma-ray flash is produced in relativistic jets of matter ejected by the explosion of the core of a very-massive star (long GRB>2 s) or the merger of compact objects such as neutron stars (short GRBs<2 s). Subsequent shock-waves produce an afterglow emission, observable in X-rays, visible, IR and radio-waves during hours after the event. The scientific field of GRB studies covers phenomenology of supernovae and mergers, physics of jets and particle acceleration, astrophysics of host galaxies, cosmological studies of the star-formation history of the Universe and fundamental physics studies through the link of GRBs with gravitational waves, neutrinos and cosmic-rays.

Onboard the future French-Chinese mission for GRB studies SVOM [1][2] (Space-based Variable Objects Monitor, Figure 1a), the wide field-of-view (FOV) soft gamma-ray telescope ECLAIRs triggers on GRB events. It detects and localizes on-board and in near real-time the appearance of short flashes of gamma-rays at random points on the sky, lasting from sub-seconds to several minutes. It requests the autonomous slew of the SVOM spacecraft within minutes, for rapid follow-up observations of the GRB afterglow emission with the onboard Microchannel X-ray Telescope (MXT) and Visible-band Telescope (VT) in order to fully characterize the event. It sends near real-time alert messages to ground via a VHF network for the Ground Follow-up Telescopes (GFTs) and Ground Wide-Angle Cameras (GWAC) of SVOM and to the whole scientific community. The observing strategy of the low-Earth-orbit satellite SVOM is such that most GRBs are detected towards the Earth-night-sky, allowing a high fraction of immediate ground-based follow-up observations, at the expense of Earth transits through the ECLAIRs FOV each orbit, which masks in mean about 30% of the observable sky.

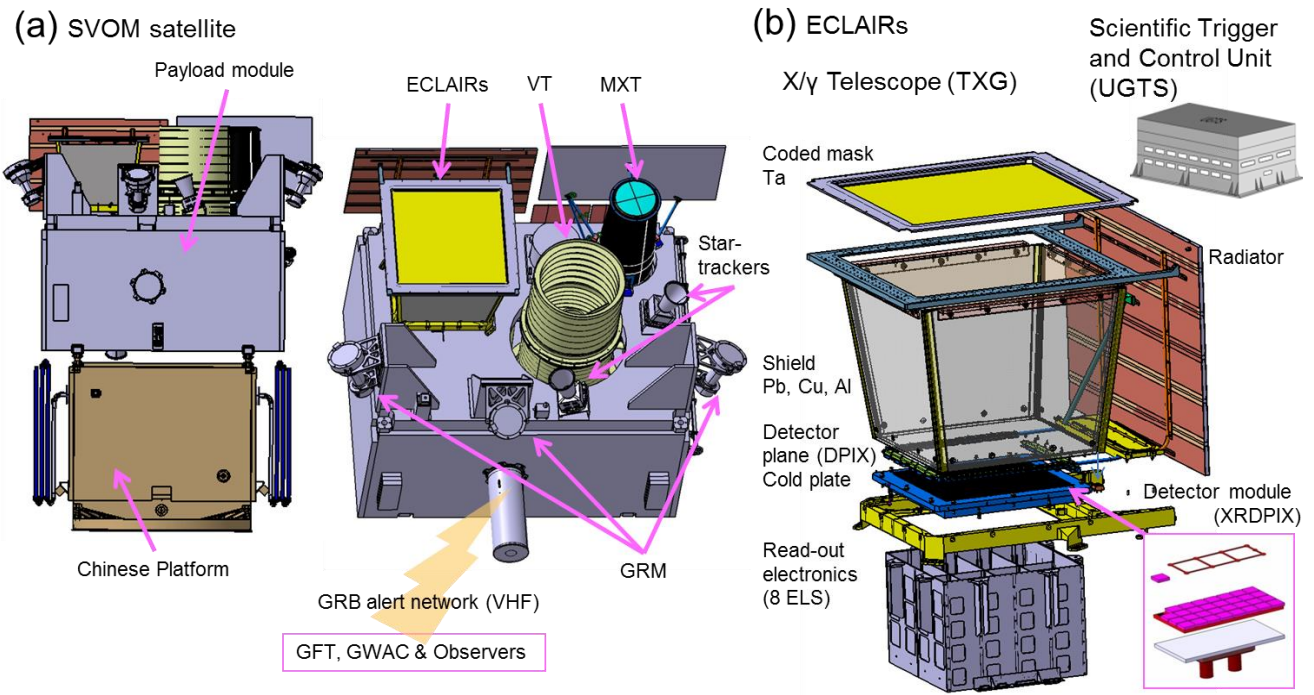

Figure 1: (a) The SVOM satellite-mission with the onboard gamma-ray imager ECLAIRs and spectrometer GRM, the onboard narrow-field instruments MXT and VT, and the ground based telescopes GFT and GWAC. (b) Schematic view of ECLAIRs composed of the X/γ Telescope and the onboard processing electronics UGTS.

## 2. The ECLAIRs telescope onbord SVOM

ECLAIRs [3], the soft gamma-ray imaging instrument of SVOM, built for autonomous onboard GRB detection and localization, is developed by a consortium of French institutes (IRAP Toulouse: PI, CEA Saclay, APC Paris, IAP Paris) and the French space agency (CNES).

ECLAIRs is formed by a coded-mask imaging telescope (TXG) with catches the randomly appearing GRBs in its large FOV (2 sr, partially coded), and the associated Scientific Trigger and Control Unit (UGTS) which performs the onboard processing and generates the GRB Triggers for SVOM. SVOM being a mini-satellite, the mass and power allocation for ECLAIRs is constrained to only 87 kg and 84 W, which permits to accommodate an active detector area of 1024 $cm^2$. Energies are detected from 4 to 150 keV. Thanks to the very-low energy threshold, ECLAIRs will detect standard GRBs as well as X-ray rich GRBs, and in particular highly redshifted GRBs. The coded mask (with open fraction of 40% and imaging capabilities up to 120 keV) provides localization accuracy better than 16 arcmin (90% C.L., at detection threshold), adapted to the FOV of the follow-up telescope VT. Two concurrent onboard-trigger algorithms search for unknown sources on the sky with durations ranging from 10 ms up to 20 min, and communicate their localizations in VHF alerts and satellite repointing requests. Camera hits are stored individually to mass-memory with an absolute timing of 10 μs.

### 2.1. The ECLAIRs detection plane

The detection plane (DPIX) [4] of the ECLAIRs TXG, built by IRAP Toulouse, is formed of 80×80 square low-leakage-current Schottky-CdTe detector pixels from Acrorad, each of 1 mm thickness and 4 mm size, placed on grid of spacing $d$=4.5 mm, in a modular design.

The DPIX is formed of 4×2 independent sectors. In a sector, 5×5 XRDPIX modules are connected with a sector readout-electronics (ELS) via connectors from Hypertac, through a cold-plate, which passively cools the detectors down to -20°C using a radiator. Each XRDPIX module (see Figure 2) connects a group of 4×8 pixels via a detector- and ASIC-ceramic to a dedicated low-noise read-out ASIC (the IDeF-X Application Specific Integrated Circuit), developed by CEA Saclay [5].

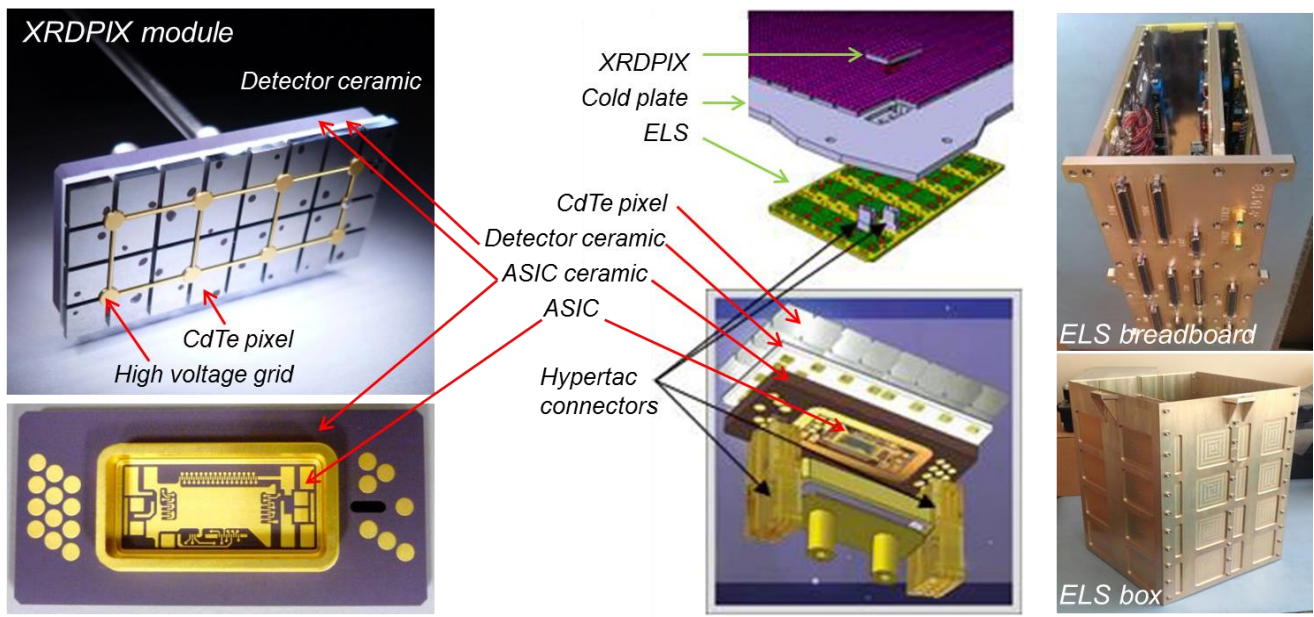

Figure 2: Photos of XRDPIX module (32 CdTe pixels assembled with an IDeF-X read-out ASIC) and ELS model.

In an ELS, ASICs with triggered channels are read-out, event times are determined (to 10 µs accuracy) and events are classified as single- or multiple-events (whether only one pixel of an XRDPIX is hit, or two or more pixels inside an XRDPIX or two or more XRDPIXs show an activity), charge digitization of single- and double-events is performed, and the pre-programmed energy calibration is applied, resulting in calibrated energy values for the events.

The 8 ELS send their events in real-time over 8 custom data-links to the UGTS, in charge of real-time data-analysis for GRB triggering, as well as data-packet building and transfer via a Spacewire-link to the Chinese payload-computer (PDPU), prior to downlink using high-bandwidth X-band telemetry, when a ground-receiver station is in view.

Over 12000 CdTe pixels have been received and tested at IRAP, 8000 have been selected as fit-for-flight, 5 proto-flight XRDPIX modules have been tested in the lab under various operating conditions [6], to validate their performance such as low-energy threshold, quantum efficiency, spectral resolution, linearity etc. Figure 3a shows an energy spectrum measured with an $^{241}$Am radioactive source in the lab, obtained by summing the calibrated spectra of the 32 pixels of an XRDPIX module (with peaking-time set to 4.4 µs in the IDeF-X ASIC), operated under -400 V bias-voltage at a temperature of -20°C in a thermal-vacuum chamber at IRAP; the low-energy threshold requirement of 4 keV is fulfilled in this test. Figure 3b shows the effective area expected from a Geant-4 Monte-Carlo simulation for the integrated detector DPIX.

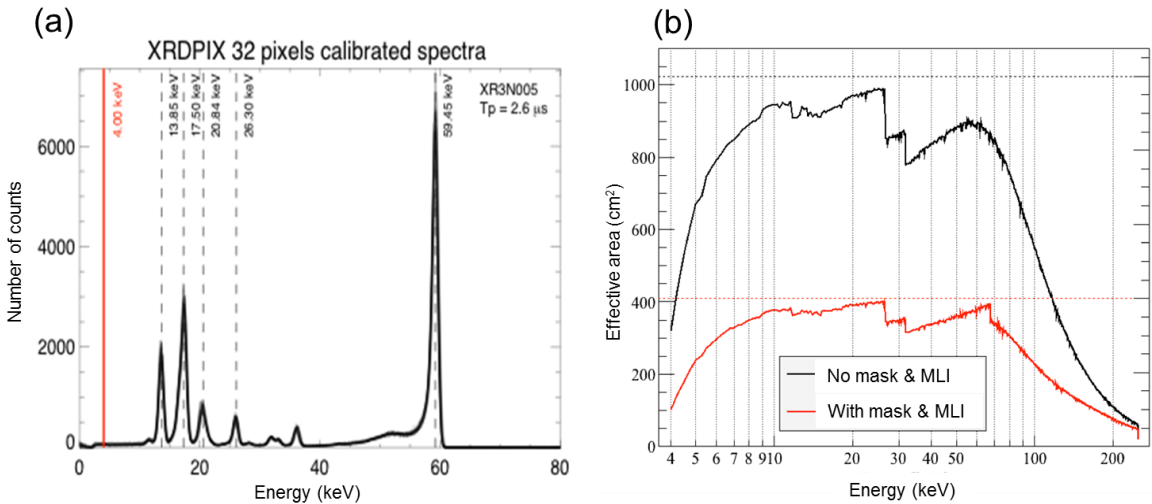

Figure 3: (a) Measured energy spectrum obtained by summing the calibrated energy spectra of the 32 pixels of an XDRPIX module, exposed to a radioactive $^{241}$Am source, confirming a low-energy threshold below 4 keV. (b) Geant-4 Monte-Carlo simulation of the effective area of the DPIX as a function of the photon energy (black curve). Adding on top of the DPIX a Ta mask of 40% open fraction and MLI (red curve) generally reduces the number of counts by this amount, except around 70 keV, where the Ta of the mask fluoresces.

## 2.2. The ECLAIRs shielding and mask

A passive graded side-shield made of Pb, Cu and Al encloses the ECLAIRs TXG (Figure 1b), to define the FOV (89°×89°) in the main energy-band (4-70 keV) by suppressing very off-axis sources and reducing the diffuse Cosmic X-ray Background (CXB), while keeping a few instrumental lines at higher energies for energy calibration.

A coded mask is placed 46 cm above the CdTe detection plane, to achieve the capability to localize sources in the FOV by deconvolution of the detector image with the mask pattern. The mask is cut-out of a Ta sheet of 0.6 mm thickness, offering an opacity which permits imaging up to 120 keV, in a self-supporting pattern of 46×46 square pixels of size *m*=11.47 mm among which 40% are empty of matter, in order to be transparent even at 4 keV (Figure 4a). For mechanical stiffness, the Ta sheet is sandwiched between two Ti foils, into which the same pattern is cut out, with holes enlarged to avoid the vignetting effects for off-axis sources (4b).

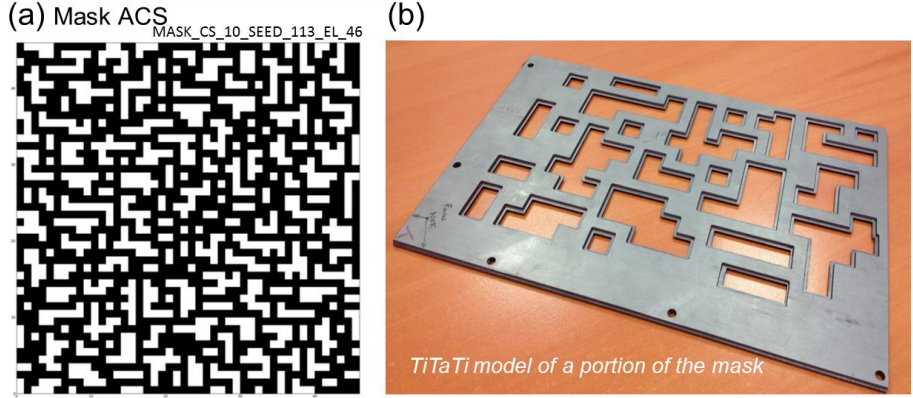

Figure 4: (a) The mask with 46×46 pixels and its pattern envisaged for flight (mask "ACS"). (b) Model of a portion of the mask, developed at APC Paris, and used for mechanical tests of the TiTaTi-sandwich concept.

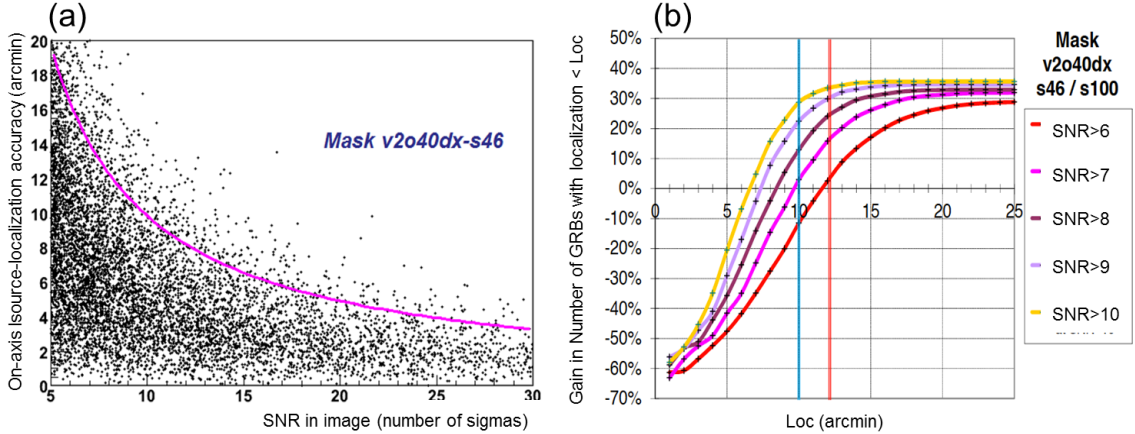

Figure 5: (a) Source localization accuracy for a test mask of 46x46 pixels (*m/d*=2.6) as a function of the signal to noise ratio (*SNRimage*) of the reconstructed sky-image peak for on-axis sources (slightly better for off-axis sources). (b) Relative gain in the expected number of detected GRBs within a given localization radius Loc and for a given *SNRimage*-threshold, between the test mask with *m/d*=2.6 compared to the previously used mask with *m/d*=1.2.

At a source detection threshold of 7σ (in *SNRimage*), such a mask with *m/d*=2.6 offers a localization accuracy of 14 arcmin at 90% CL (Figure 5a), which is degraded compared to the previously used mask with *m/d*=1.2 (8 arcmin). But the mask pattern used for deconvolution (physical pattern rebinned onto the detector grid) offers a much better contrast, such that about 30% more GRBs are expected to be detected with a mask with *m/d*=2.6 compared to 1.2, as obtained at CEA in static simulations of imaging in fluence mode using a synthetic table of realistic GRBs provided by IAP Paris, including very weak ones (Figure 5b). All detected GRBs will be localized well enough to be inside the FOV of MXT (~1°) after slew. Even for VT, with a FOV of 26 arcmin in diameter, the new mask adds a net gain of more than 15% of GRBs.

**2.3. The ECLAIRs onboard Scientific Trigger and Control Unit (UGTS)**

The UGTS, developed by CEA Saclay, performs the ECLAIRs configuration, house-keeping, temperature control management of the 8 ELS, data acquisition and transfer to the PDPU mass-memory, noisy-pixel management and power supply function for ECLAIRs, additionally to the real-time data-processing for the GRB trigger. The UGTS combines the former Scientific Trigger Unit (UTS) with two former electronic boxes (UGD and part of FCU) into a single unit, for mass savings and to allow a single Spacewire-interface with the PDPU.

Two simultaneous GRB triggers are studied for UGTS: (1) the *Count-rate Trigger* operating on 10 ms to 20 s time-scales, 4 energy bands and 9 detector zones, which detects count-rate increases over a background model and performs every 2.5 s sky imaging by mask deconvolution of the best excess found. New sources with *SNRimage* above threshold are searched and localized on the sky by peak-fitting. (2) the *Image Trigger* which performs every 20 s systematic (background corrected) sky imaging in 4 energy bands and builds by summation sky images of durations up to 20 min, in which new sources are also searched and localized.

The trigger algorithms have been coded and scientifically tested on linux machines [7] and benchmarked on target hardware [8]. Engineering-boards for the UTS, based on Atmel FPGAs and an Atmel Leon2 processor (radiation hard, space qualified and ITAR-free), have been built and tested in the lab (Figure 6). Detailed scientific performance-studies of the algorithms with simulated GRB databases are ongoing [9]. An upgrade is under study with a Leon3 dual-core processor, offering more CPU power for the wider functionalities of UGTS compared to UTS.

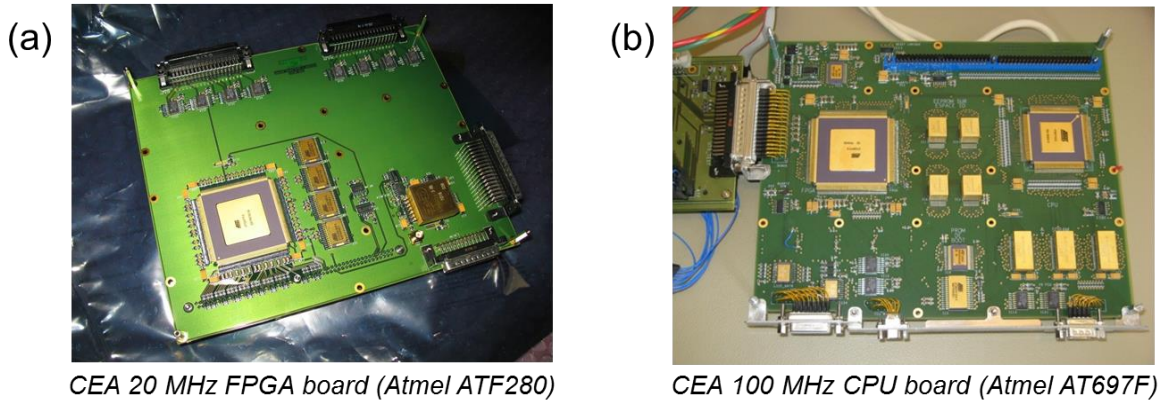

(a) CEA 20 MHz FPGA board (Atmel ATF280)     (b) CEA 100 MHz CPU board (Atmel AT697F)

Figure 6: Custom engineering boards (FPGA and Leon2-CPU board) of the former UTS, developed by CEA Saclay.

### 3. Conclusions

The SVOM mission for GRB studies has been recently approved for implementation by France and China, the kick-off of the mission Phase B took place in 2014. The developments of the ECLAIRs detector and electronics are on track for a launch currently scheduled in 2021. We expect to detect about 80 GRBs/yr with ECLAIRs, including (thanks to its low-energy threshold) about 34% of X-ray rich GRBs, among which many high-redshifted GRBs, well localized to be quickly followed-up, onboard by MXT and VT, and by the whole community.